\documentstyle[epsfig,preprint,aps]{revtex}
\begin{document}
\draft
\tighten
\preprint{\vbox{
\hbox{CLNS 97/1530 \hfill}
\hbox{CLEO 97-29 \hfill}
\vskip1cm
}}

\title{New Limits for Neutrinoless Tau Decays}

\date{\today}
\maketitle

\begin{abstract}
Neutrinoless 3-prong tau lepton decays into a charged lepton
and either two charged particles or one neutral meson
have been searched for using 4.79~fb$^{-1}$ of data collected
with the CLEO~II detector at CESR.  This analysis represents an update
of a previous study and the addition of six decay
channels. In all channels the numbers of events
found are compatible with background estimates and
branching fraction upper limits are set
for 28 different decay modes. These limits are either more stringent
than those set previously or represent the first attempt to find these
decays.
\end{abstract}


\begin{center}
D.~W.~Bliss,$^{1}$ G.~Masek,$^{1}$ H.~P.~Paar,$^{1}$
S.~Prell,$^{1}$ V.~Sharma,$^{1}$
D.~M.~Asner,$^{2}$ J.~Gronberg,$^{2}$ T.~S.~Hill,$^{2}$
D.~J.~Lange,$^{2}$ R.~J.~Morrison,$^{2}$
% ...
\end{center}
 
\small
\begin{center}
$^{1}${University of California, San Diego, La Jolla, California 92093}\\
$^{2}${University of California, Santa Barbara, California 93106}\\
% ...
$^{13}${University of Minnesota, Minneapolis, Minnesota 55455}\\
$^{14}${State University of New York at Albany, Albany, New York 12222}\\
% ...
\end{center}

\newpage
In the Standard Model of electroweak interactions the difference
between the number of leptons and the number of antileptons is
conserved for each generation separately. However, there is no
fundamental motivation for this lepton flavor conservation in this theory
because there is no symmetry associated with lepton family
number. Many extensions of the Standard Model predict flavor
violation in lepton decays. Among them are
models that involve heavy neutral
leptons~\cite{ref:gonz,ref:wong,ref:pilaf,ref:hisano,ref:ilackpil,ref:ilackkniehl,ref:ilack},
left-right symmetries~\cite{ref:Maho,ref:moha,ref:leftr},
supersymmetry~\cite{ref:susy,ref:bhat,ref:choud,ref:rpar} or
superstrings~\cite{ref:kelley,ref:Arno,ref:jwu}.
The expected decay branching fractions in these models
depend on the unknown masses of proposed new particles and
on the new coupling constants. The most optimistic branching
fraction predictions are at the level of about $10^{-6}$.
Constraints on lepton flavor violation 
come from studies of rare and forbidden $K$, $\pi$, and $\mu$ decays,
$e$-$\mu$ conversions, neutrinoless double beta decays, neutrino oscillations,
$Z\rightarrow l_{1}^{+}l_{2}^{-}$ decays, and
other rare processes.
In particular, there are strict limits on muon neutrinoless
decays: $B(\mu\rightarrow e\gamma) < 4.9\times 10^{-11}$ and
$B(\mu\rightarrow eee) < 2.4\times 10^{-12}$ at 90\% confidence
level~\cite{ref:PDG}.
However, lepton number violation rates may exhibit a strong dependence
on mass and on generation number of the decaying particle,
thus enhancing tau lepton decay rates.
Also, the larger mass of the tau allows for new
decay types which are kinematically forbidden for the muon.

The CLEO collaboration has already performed comprehensive searches
for neutrinoless tau decays in various
channels~\cite{ref:igv,ref:narsky,ref:jiklee}. The analysis presented
in this paper updates the results of Ref.~\cite{ref:igv} with a more
than twofold increase in the dataset size. The search also
includes six additional channels. 
A detailed description of this analysis can be
found in Ref.~\cite{ref:thesis}. We search for tau
decays into three charged particles:
\begin{eqnarray*}
&&\tau^{\pm}\rightarrow(l_{1}l_{2}l_{3})^{\pm},\ \ \ l = e~\mbox{or}~\mu,\\
&&\tau^{\pm}\rightarrow(lh_{1}h_{2})^{\pm},\ \ \ h = 
\pi^{\pm}~\mbox{or}~K^{\pm}.
\end{eqnarray*}
All possible combinations of final state particles and charge assignments
are considered, except for those
that do not conserve electric charge.
Different assignments result in either lepton flavor
violating decays,
as in $\tau^{-}\rightarrow\mu^{-}e^{+}e^{-}$, or both
lepton flavor and lepton number violating decays, as in
$\tau^{-}\rightarrow e^{+}\pi^{-}\pi^{-}$.
We also search for $\tau$ decays into one charged
lepton and one neutral meson which can 
subsequently decay into two charged hadrons
thus resulting in three charged particles
in the final state:
\[ \tau^{\pm}\rightarrow l^{\pm}M,\ \ \ M =
\rho^{0},\,\phi,\,K^{*0},~\mbox{or}~\bar K^{*0}. \]
The channels with two charged kaons, possibly coming
from the decay of $\phi$ meson, have been searched for the first time.

The data used in this analysis were collected with the CLEO~II
detector~\cite{ref:detector} at the Cornell Electron Storage Ring (CESR). 
We use information from a 67-layer
tracking system which also provides specific ionization measurements
(dE/dx), time-of-flight scintillation counters
and a 7800-crystal CsI calorimeter.
These elements are inside a 1.5\,T superconducting
solenoidal magnet whose iron yoke also serves
as a hadron absorber for a muon identification system.
Tau leptons were produced in pairs in $e^+e^-$ collisions
at a center-of-mass energy
of about 10.6~GeV. The data correspond to an integrated luminosity
of 4.79~fb$^{-1}$, and the number of produced tau pairs is $4.37\cdot10^6$. 

We follow the search method described in Ref.~\cite{ref:igv}.
Signal candidate tau decays are required to produce three well-reconstructed
tracks in the detector (3-prong decay). The other tau in the event must
decay into a 1-prong mode, and the total visible charge
must be zero. Not more than one photon candidate or background
shower in the CsI calorimeter is allowed
on the 3-prong side of the event. 

At least one charged particle on the 3-prong side is required to satisfy
electron or muon identification criteria. These criteria are more
strict in the $\tau^{\pm}\rightarrow(lh_{1}h_{2})^{\pm}$
and the $\tau^{\pm}\rightarrow l^{\pm}M$ decay channels than in the
$\tau^{\pm}\rightarrow(l_{1}l_{2}l_{3})^{\pm}$
channels because of large background from tau decays into three pions
and a neutrino, in which one of the pions is misidentified as an electron
or a muon. Electrons are identified by requiring that the ratio of
the energy deposited by the particle 
in the CsI calorimeter to the momentum measured in the
drift chamber is close to unity. Muon candidates are required
to have a well-reconstructed track in the
muon system. Charged mesons ($\pi$ or $K$) are not positively
identified, and we try all possible meson type assignments
to tracks. Thus, a single event can be a candidate for
more than one final state.
In the channels involving neutral
mesons we require the two-hadron invariant mass to be 
consistent with that of the corresponding meson: 
$M_{\pi^{+} \pi^{-}} < 1.2~\mbox{GeV/c}^{2}$ for $\rho^{0}$, 
$0.7~\mbox{GeV/c}^{2} < M_{\pi^{\pm} K^{\mp}} < 1.1~\mbox{GeV/c}^{2}$
for $K^{*\,0}$ and $\bar K^{*\,0}$,
$0.99~\mbox{GeV/c}^{2} < M_{K^{\pm} K^{\mp}} < 1.05~\mbox{GeV/c}^{2}$
for $\phi$, where the mass interval is based on the resonance 
width and the detector resolution. 

The main backgrounds remaining after application of particle
identification requirements are photon conversions
in radiative Bhabha and muon pair events, two-photon
processes, low multiplicity hadronic events, and 
$\tau\rightarrow 3h\nu_{\tau}$ decays in which 
at least one hadron is misidentified as a lepton.
Photon conversions in the detector material
produce $e^{+}e^{-}$ pairs with small invariant masses.
To suppress conversions, we consider each pair of oppositely
charged particles not
identified as muons under the assumption that
both particles are electrons. Events are rejected if the
invariant mass is less than 0.15~GeV/c$^{2}$ for any such pair.
Two-photon processes have low values of the
total transverse momentum with respect
to the beam direction. In contrast, the
signal events have at least one undetected neutrino on the
1-prong side which leads to transverse momentum imbalance.
We reduce two-photon background contribution
by requiring transverse momentum in excess of 0.2~GeV/c.
Neutrino presence in the event
is further exploited by requiring at least 3$^{\circ}$
acollinearity between the direction of the sum of charged particles'
momenta on the 3-prong side of the event and the direction
of the 1-prong momentum.
For neutrinoless decays the sum of the four-momenta of the 
particles on the 3-prong side define the tau direction and energy.
Neglecting radiative effects, the other tau in the event has an
opposite momentum vector. We determine the momentum
of the 1-prong charged particle in the rest system of a
parent tau with boost parameters obtained by 
summing the four-momenta of the 3-prong side particles.
Momentum values larger than half of tau mass are
kinematically forbidden for tau decay
products in the rest frame of decaying particle. 
However, presence of a neutrino on the 3-prong
side of the event may result in incorrect determination of boost
parameters and higher momentum. We require that the 1-prong momentum
in the parent tau rest frame is less than 1~GeV/c,
thus reducing background from standard tau decay modes.

The efficiencies of the selection criteria were estimated using
16\,000 Monte Carlo events for each decay channel. Phase
space distributions were used to generate neutrinoless tau decays in
all the channels.  The KORALB/TAUOLA program package~\cite{ref:koral}
was used to simulate the tau-pair production and the decay of the
1-prong tau. Subsequent meson decays and decays of the 1-prong tau were
generated according to the known branching
fractions~\cite{ref:PDG}. Detector signals were simulated with the
GEANT-based CLEO~II simulation program~\cite{ref:soft}.

For neutrinoless tau decays
the total energy measured on the 3-prong side, $E_{3}$, must 
be equal to the beam energy, $E_{\text{beam}}$, and the invariant mass
of the three charged particles, $M_{3}$,
must be equal to the tau mass. 
For all channels we select rectangular signal regions
in the $E_{3}-E_{\text{beam}}$ and $M_{3}$ variables
taking into account detector
resolution, signal efficiencies, and background levels.
The signal region optimization algorithm (minimization
of average expected upper limits) is described in detail
in Ref.~\cite{ref:thesis}. We assign the channels
studied to three different groups according to their
background density. For the channels with low
background ($\tau^{-}\rightarrow e^{-}e^{+}e^{-},\  e^{-}\phi,
\ e^{+}\mu^{-}\mu^{-},\ \mbox{and}\ \mu^{+}e^{-}e^{-},$
where charge conjugated
modes are  always implied) we define the signal region as
\begin{eqnarray*}
&-0.39 \ \mbox{GeV}  < E_{3} - E_{beam} <  0.08 \ \mbox{GeV},&\\
&1.70 \ \mbox{GeV/c}^2  < M_{3} <  1.81 \ \mbox{GeV/c}^2.&
\end{eqnarray*}
For the medium background channels 
($\tau^{-}\rightarrow e^{-}\mu^{+}\mu^{-},$ $\mu^{-}\mu^{+}\mu^{-},$
$e^{+}\pi^{-}K^{-},$ $\mu^{-}e^{+}e^{-},$ $e^{+}\pi^{-}\pi^{-},$
$e^{+}K^{-}K^{-},$ and $\mu^{-}\phi$) we require
\begin{eqnarray*}
&-0.17 \ \mbox{GeV}  < E_{3} - E_{beam} <  0.09 \ \mbox{GeV},&\\
&1.74 \ \mbox{GeV/c}^2  < M_{3} <  1.80 \ \mbox{GeV/c}^2,&
\end{eqnarray*}
and for the rest of the channels (high background group) we require
\begin{eqnarray*}
&-0.09 \ \mbox{GeV}  < E_{3} - E_{beam} <  0.06 \ \mbox{GeV},&\\
&1.75 \ \mbox{GeV/c}^2  < M_{3} <  1.80 \ \mbox{GeV/c}^2.&
\end{eqnarray*}

\begin{figure}[p]
\begin{center}
\scalebox{1.0}{\includegraphics{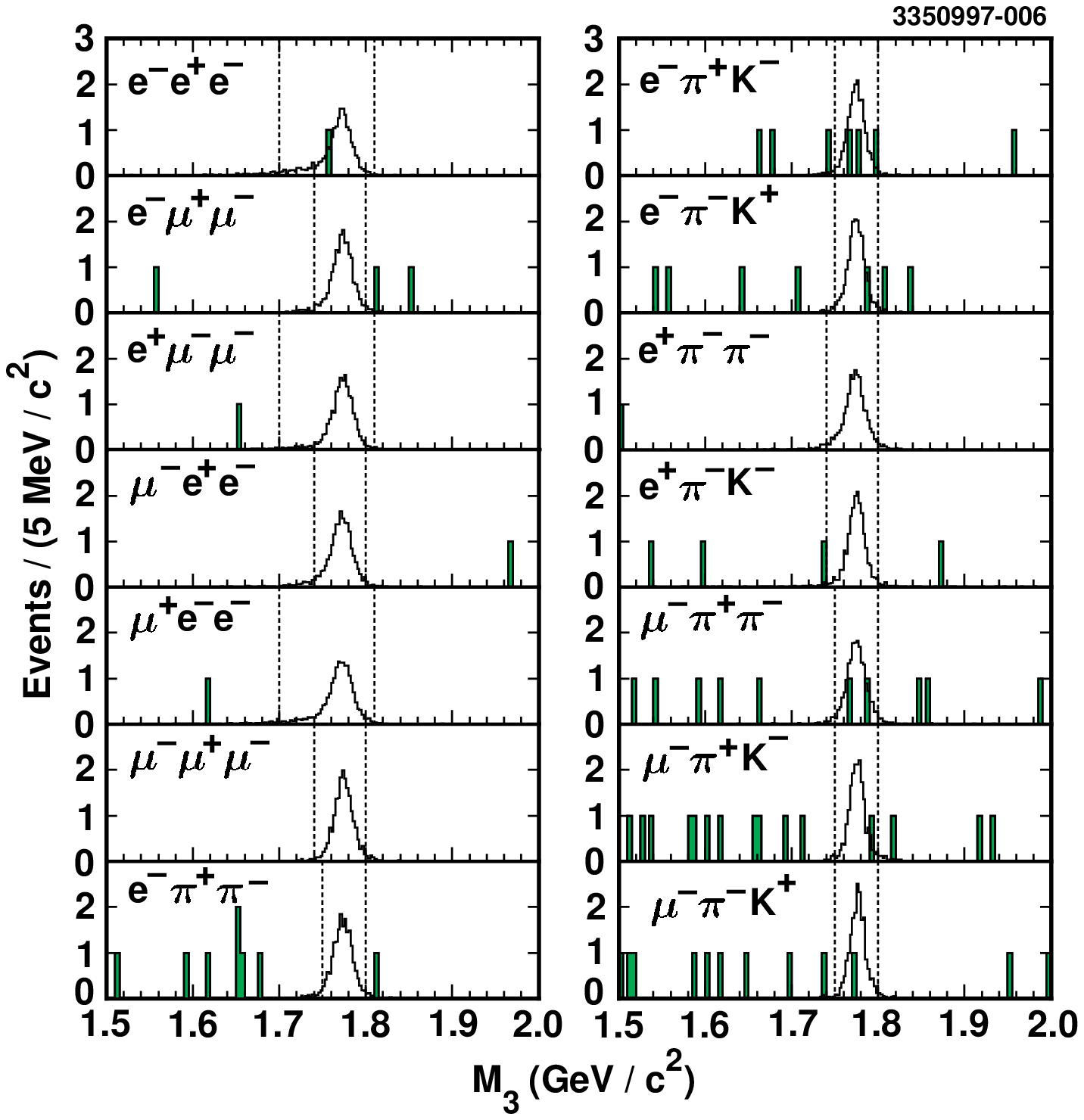}}
\end{center}
\caption{\label{fig:mdistr1}
Distributions of the invariant mass of the 3-prong side particles,
$M_{3}$, for the data (shaded histogram) and signal Monte Carlo events
(solid line).
The expected signal shapes are shown with arbitrary
normalization. The dotted lines indicate the boundaries of the
signal regions used. See also Fig.~\protect\ref{fig:mdistr2}.
}
\end{figure}

\begin{figure}[p]
\begin{center}
\scalebox{1.0}{\includegraphics{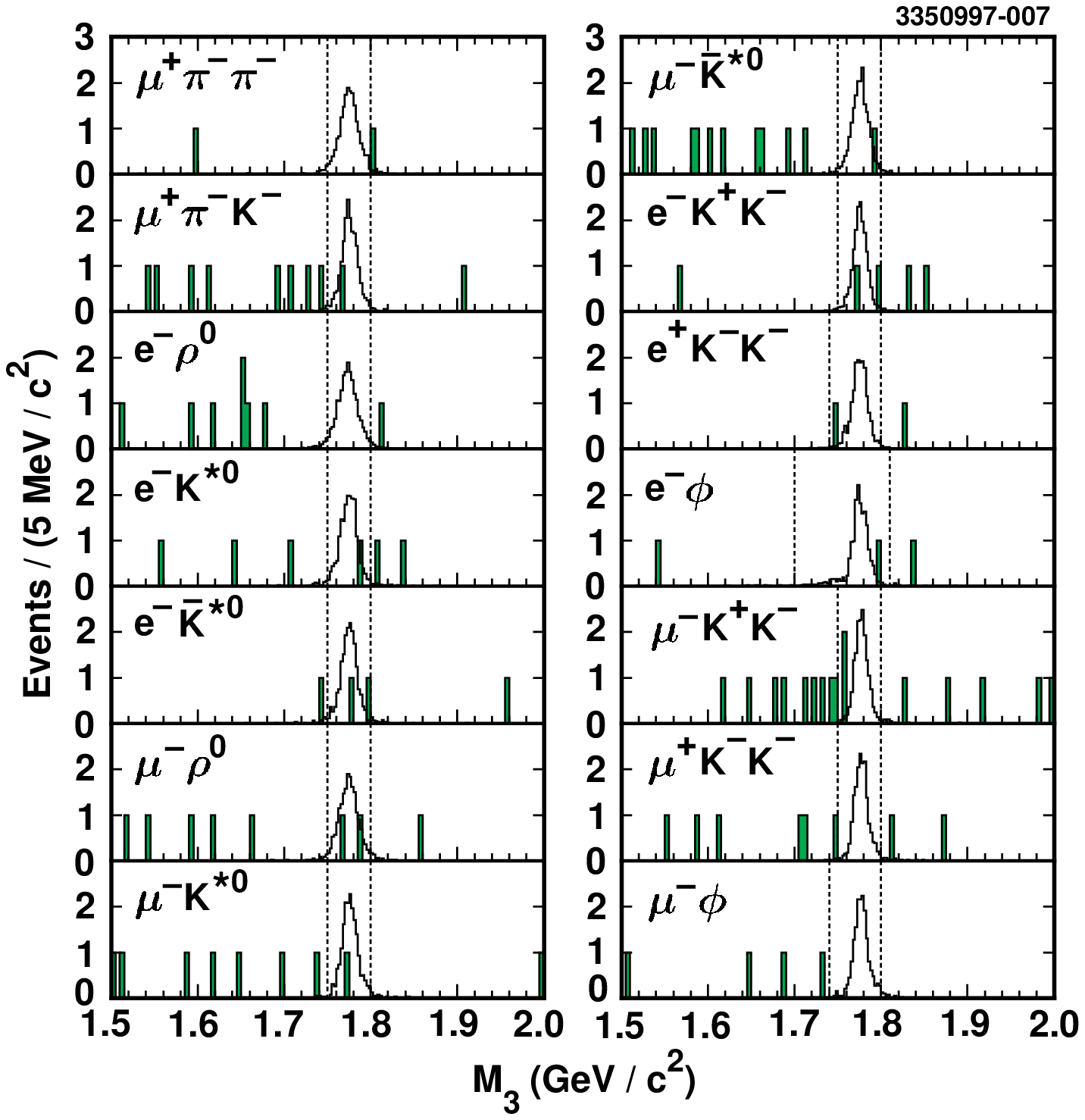}}
\end{center}
\caption{\label{fig:mdistr2}
Distributions of the invariant mass of the 3-prong side particles,
$M_{3}$, for the data (shaded histogram) and signal Monte Carlo events
(solid line).
The expected signal shapes are shown with arbitrary
normalization. The dotted lines indicate the boundaries of the
signal regions used. See also Fig.~\protect\ref{fig:mdistr1}.
}
\end{figure}

The 3-prong invariant mass distributions of events satisfying all
background suppression criteria and lying within the $E_{3}-E_{beam}$
limits defined above are shown in Figs.~\ref{fig:mdistr1} and
~\ref{fig:mdistr2}, together with the expected signal shapes generated
by the Monte Carlo simulation.  There are 14 events in the data which
satisfy all the selection criteria, including the 3-prong invariant
mass requirement, in at least one channel. 7 of these events satisfy
the selection criteria in two different channels (notably, most
candidates for $\tau^{\pm}\rightarrow l^{\pm}M$ decays also qualify
for corresponding non-resonant decays), and 1 event satisfies the
selection criteria in three different channels. In each channel the
number of data events inside the signal region is consistent with the
estimated background level. The largest deviation is observed in the
$\tau^{-}\rightarrow e^{-}\pi^{+}K^{-}$ channel which has three events
while the expected background is 0.42 events. The probability of such
a deviation or larger is about 1\%, if calculated
according to Poisson statistics.  However, with 28 channels
investigated, such fluctuations can be expected in one or two of them.
In addition, Poisson statistics may fail to provide an accurate
consistency check because, due to the small size of the event sample
remaining after background suppression, we use the same sideband
data for both signal region optimization and background estimation.

In each channel we calculate the branching
fraction upper limit at the 90\% confidence level
according to the convention adopted by
the Particle Data Group~\cite{ref:PDG},
and we do not attempt to subtract the background.
Systematic errors in this analysis arise from
uncertainties in our knowledge of the luminosity,
track reconstruction efficiency,
lepton identification efficiency, and 3-prong
energy and invariant mass resolutions.
Combined together, they are conservatively
estimated to increase branching fraction upper limits by 10\%.
We do not assign any systematic error due to model dependence.
However, we emphasize that our limits depend on the assumed angular
and momentum distributions of the decay particles. 

The final results are summarized in Table~\ref{tab:results}, together
with the detection efficiencies obtained for each mode from Monte
Carlo simulations and with the numbers of events observed in the data.
The limits obtained in this analysis are more stringent than those
obtained previously~\cite{ref:PDG}. In addition, the limits on
$B(\tau^{-}\rightarrow \mu^{+} e^{-}e^{-})$ and $B(\tau^{-}\rightarrow
e^{+}\mu^{-}\mu^{-})$ are the most stringent limits to date on lepton
number violation in $\tau$ decays.
In SUSY with broken R-parity~\cite{ref:choud,ref:rpar}
and in the model with radiatively generated
lepton masses from Ref.~\cite{ref:wong}
the obtained results provide constraints on model
parameters. In models with heavy
neutral leptons~\cite{ref:ilackpil,ref:ilackkniehl,ref:ilack}
the experimental limits are close to the allowed range of
neutrinoless $\tau$ decay rates.

We gratefully acknowledge the effort of the CESR staff in providing us with
excellent luminosity and running conditions.
J.P.A., J.R.P., and I.P.J.S. thank                                           
the NYI program of the NSF, 
M.S. thanks the PFF program of the NSF,
G.E. thanks the Heisenberg Foundation, 
K.K.G., M.S., H.N.N., T.S., and H.Y. thank the
OJI program of DOE, 
J.R.P., K.H., M.S. and V.S. thank the A.P. Sloan Foundation,
R.W. thanks the 
Alexander von Humboldt Stiftung,
M.S. thanks Research Corporation, 
and S.D. thanks the Swiss National Science Foundation 
for support.
This work was supported by the National Science Foundation, the
U.S. Department of Energy, and the Natural Sciences and Engineering Research 
Council of Canada.

\begin{table}[p]
\caption{\label{tab:results}
Detection efficiencies, event statistics, expected backgrounds,
and upper limits for branching fractions at~90\%~confidence level.
}
\begin{center}
\begin{tabular}{lcccc}
Decay   & Detection      & Events   & Expected  & Upper \\
channel & efficiency, \% & observed & bg events & limits, $10^{-6}$ \\
\tableline
$\tau^{-}\rightarrow e^{-}e^{+}e^{-}$
& 17.0 & 1 & 0.21 & 2.9 \\
$\tau^{-}\rightarrow \mu^{-} e^{+}e^{-}$
& 16.8 & 0 & 0.18 & 1.7 \\
$\tau^{-}\rightarrow \mu^{+} e^{-}e^{-}$
& 19.5 & 0 & 0.12 & 1.5 \\
$\tau^{-}\rightarrow e^{-}\mu^{+}\mu^{-}$ 
& 16.5 & 0 & 0.32 & 1.8 \\
$\tau^{-}\rightarrow e^{+}\mu^{-}\mu^{-}$ 
& 19.9 & 0 & 0.12 & 1.5 \\
$\tau^{-}\rightarrow \mu^{-}\mu^{+}\mu^{-} $ 
& 15.0 & 0 & 0.11 & 1.9 \\
$\tau^{-}\rightarrow e^{-}\pi^{+}\pi^{-} $ 
& 13.2 & 0 & 0.43 & 2.2 \\
$\tau^{-}\rightarrow e^{-}\pi^{-}K^{+} $ 
& 13.0 & 1 & 0.29 & 3.8 \\
$\tau^{-}\rightarrow e^{-}\pi^{+}K^{-} $ 
& 13.1 & 3 & 0.42 & 6.4 \\
$\tau^{-}\rightarrow e^{-}K^{+}K^{-} $
& 11.2 & 2 & 0.29 & 6.0 \\
$\tau^{-}\rightarrow e^{+}\pi^{-}\pi^{-} $ 
& 15.3 & 0 & 0.22 & 1.9 \\
$\tau^{-}\rightarrow e^{+}\pi^{-}K^{-} $ 
& 14.0 & 0 & 0.18 & 2.1 \\
$\tau^{-}\rightarrow e^{+}K^{-}K^{-} $
& 13.0 & 1 & 0.11 & 3.8 \\
$\tau^{-}\rightarrow \mu^{-}\pi^{+}\pi^{-} $ 
& 8.2 & 2 & 0.57 & 8.2 \\
$\tau^{-}\rightarrow \mu^{-}\pi^{-}K^{+} $ 
& 6.7 & 1 & 0.48 & 7.4 \\
$\tau^{-}\rightarrow \mu^{-}\pi^{+}K^{-} $ 
& 6.5 & 1 & 0.49 & 7.5 \\
$\tau^{-}\rightarrow \mu^{-}K^{+}K^{-} $
& 4.5 & 2 & 0.50 & 15 \\
$\tau^{-}\rightarrow \mu^{+}\pi^{-}\pi^{-} $ 
& 8.6 & 0 & 0.36 & 3.4 \\
$\tau^{-}\rightarrow \mu^{+}\pi^{-}K^{-} $ 
& 7.0 & 1 & 0.33 & 7.0 \\
$\tau^{-}\rightarrow \mu^{+}K^{-}K^{-} $
& 4.8 & 0 & 0.35 & 6.0 \\
$\tau^{-}\rightarrow e^{-}\rho^{0} $
& 14.4 & 0 & 0.45 & 2.0 \\
$\tau^{-}\rightarrow e^{-} K^{*\,0} $
& 9.5 & 1 & 0.32 & 5.1 \\
$\tau^{-}\rightarrow e^{-} \bar{K}^{*\,0} $
& 9.0 & 2 & 0.32 & 7.4 \\
$\tau^{-}\rightarrow e^{-}\phi $
& 7.2 & 1 & 0.15 & 6.9 \\
$\tau^{-}\rightarrow \mu^{-}\rho^{0} $
& 10.6 & 2 & 0.43 & 6.3 \\
$\tau^{-}\rightarrow \mu^{-} K^{*\,0} $
& 6.5 & 1 & 0.46 & 7.5 \\
$\tau^{-}\rightarrow \mu^{-} \bar{K}^{*\,0} $
& 6.5 & 1 & 0.37 & 7.5 \\
$\tau^{-}\rightarrow \mu^{-}\phi $
& 4.1 & 0 & 0.11 & 7.0 \\
\end{tabular}
\end{center}
\end{table}

\newpage

\begin{thebibliography}{99}

\bibitem{ref:gonz}{M.C. Gonzalez-Garsia and J.W.F. Valle,
Mod. Phys. Lett. {\bf 7}, 477 (1992).}

\bibitem{ref:wong}{G.-G. Wong and W.-S. Hou,
Phys. Rev. D {\bf 50}, 2962 (1994).}

\bibitem{ref:pilaf}{A. Pilaftsis,
Mod. Phys. Lett. A {\bf 9}, 3595 (1994).}

\bibitem{ref:hisano}{J. Hisano {\it et al.},
Phys. Lett. B {\bf 357}, 579 (1995).}

\bibitem{ref:ilackpil}{A. Ilakovac and A. Pilaftsis,
Nucl. Phys. B {\bf 437}, 491 (1995).}

\bibitem{ref:ilackkniehl}{A. Ilakovac, B.A. Kniehl, and A.Pilaftsis,
Phys. Rev. D {\bf 52}, 3993 (1995).}

\bibitem{ref:ilack}{A. Ilakovac,
Phys. Rev. D {\bf 54}, 5653 (1996).}

\bibitem{ref:Maho}{R.N. Mohapatra,
Phys. Rev. D {\bf 46}, 2990 (1992).}

\bibitem{ref:moha}{R.N. Mohapatra, S. Nussinov and X. Zhang,
Phys. Rev. D {\bf 49}, 2410 (1994).}

\bibitem{ref:leftr}{S. Pastor, S.D. Rindani, and J.W.F. Valle,
FTUV-97-23, hep-ph/9705394 (1997).}

\bibitem{ref:susy}{J.C. Romao, N. Rius, and J.W.F. Valle,
Nucl. Phys. B {\bf 363}, 369 (1991).}

\bibitem{ref:bhat}{G. Bhattacharyya and D. Choudhury,
Mod. Phys. Lett. A {\bf 10}, 1699 (1995).}

\bibitem{ref:choud}{D. Choudhury and P. Roy,
Phys. Lett. B {\bf 378}, 153 (1996).}

\bibitem{ref:rpar} {J.E. Kim, P. Ko, D.-G. Lee,
Phys. Rev. D {\bf 56}, 100, (1997).}

\bibitem{ref:kelley}{S. Kelley {\it et al.},
Nucl. Phys. B {\bf 358}, 27 (1991).}

\bibitem{ref:Arno}{R. Arnowitt and P. Nath,
Phys. Rev. Lett. {\bf 66}, 2708 (1991).}

\bibitem{ref:jwu}{J. Wu, S. Urano, and R. Arnowitt,
Phys. Rev. D {\bf 47}, 4006 (1993).}

\bibitem{ref:PDG} { Particle Data Group, R.M. Barnett {\it et al.},
Phys. Rev. {\bf D54}, Part I (1996).}

\bibitem{ref:igv}{J.~Bartelt {\it et al.}, 
Phys. Rev. Lett. {\bf 73}, 1890 (1994).}

\bibitem{ref:narsky}{K.~Edwards {\it et al.},
Phys. Rev. D {\bf 55}, 3919 (1997).}

\bibitem{ref:jiklee}{G.~Bonvicini {\it et al.},
Phys. Rev. Lett. {\bf 79}, 1221 (1997).}

\bibitem{ref:thesis}{I.~Volobouev,
``Rare and Forbidden Decays of the Tau Lepton'', 
Ph.D. thesis, Southern Methodist University, SMUPHT/97-2 (1997).}

\bibitem{ref:detector} {Y. Kubota {\it et al.},
Nucl. Instrum. Meth. {\bf A320}, 66 (1992).}

\bibitem{ref:koral}{S. Jadach and Z. W{\c{a}}s,
Comput. Phys. Commun. {\bf 64}, 267 (1991);
S. Jadach {\it et al.}, Comput. Phys. Commun. {\bf 76}, 361 (1993).}

\bibitem{ref:soft} {The CLEO~II detector simulation
is based on the GEANT software
package: R. Brun {\it et. al.}, GEANT version 3.15, CERN DD/EE/84-1.}

\end{thebibliography}
\end{document}